\documentclass{article}
\usepackage[utf8]{inputenc}

\usepackage{hyperref}
\usepackage[sorting=none,style=ieee]{biblatex}
\usepackage{amsfonts}
\usepackage{amsmath}
\usepackage{graphicx}
\usepackage{algorithm}
\usepackage{algpseudocode}
\usepackage{mathtools}
\usepackage{textalpha}
\usepackage[title]{appendix}

\newcommand{\Tau}{\mathrm{T}}
\newcommand{\diag}{\mathrm{diag}}
\newcommand{\trho}{T$_{1\rho}$} % Version for text
\newcommand{\trhoTwo}{\mathrm{T_{1\rho}}} % Version for equations
\newcommand{\szero}{S$_0$} % Version for text
\newcommand{\szeroTwo}{\mathrm{S_0}} % Version for equations
\DeclarePairedDelimiter\abs{\lvert}{\rvert}%
\DeclarePairedDelimiter\norm{\lVert}{\rVert}%
\DeclareMathOperator*{\argmin}{arg\,min}
\makeatletter
\let\oldabs\abs
\def\abs{\@ifstar{\oldabs}{\oldabs*}}
\let\oldnorm\norm
\def\norm{\@ifstar{\oldnorm}{\oldnorm*}}
\makeatother

\title{Embedded quantitative MRI \trho{} mapping using non-linear primal-dual proximal splitting}
\author{Matti Hanhela$^1$\thanks{matti.hanhela@uef.fi} \and
        Antti Paajanen$^1$ \and
        Mikko J. Nissi$^1$ \and
        Ville Kolehmainen$^1$ \\
        \small$^1$ Department of Applied Physics, University of Eastern Finland}
\date{\today}

\begin{document}

\maketitle

%%%%%%%%%%%%%%%%%%%%%%%%%%%%%%%%%%%%%%%%%%%%%%%%%%%%%%%%%%%%%%%%%%
\begin{abstract}
Quantitative MRI (qMRI) methods allow reducing the subjectivity of clinical MRI by providing numerical values on which diagnostic assessment or predictions of tissue properties can be based. However, qMRI measurements typically take more time than anatomical imaging due to requiring multiple measurements with varying contrasts for, e.g., relaxation time mapping. To reduce the scanning time, undersampled data may be combined with compressed sensing reconstruction techniques. Typical CS reconstructions first reconstruct a complex-valued set of images corresponding to the varying contrasts, followed by a non-linear signal model fit to obtain the parameter maps. We propose a direct, embedded reconstruction method for \trho{} mapping. The proposed method capitalizes on a known signal model to directly reconstruct the desired parameter map using a non-linear optimization model. The proposed reconstruction method also allows directly regularizing the parameter map of interest, and greatly reduces the number of unknowns in the reconstruction. We test the proposed model using a simulated radially sampled data from a 2D phantom and 2D cartesian \emph{ex vivo} measurements of a mouse kidney specimen. We compare the embedded reconstruction model to two CS reconstruction models, and in the cartesian test case also iFFT. The proposed, embedded model outperformed the reference methods on both test cases, especially with higher acceleration factors.
\end{abstract}

%%%%%%%%%%%%%%%%%%%%%%%%%%%%%%%%%%%%%%%%%%%%%%%%%%%%%%%%%%%%%%%%%%
\section{Introduction}
Magnetic resonance imaging (MRI) is one of the most important tools for clinical diagnosis of various diseases, due to its excellent and versatile soft tissue contrast. Clinical MRI is based on expert interpretation of anatomical images of varying contrasts and thus tends to retain a level of subjectivity. Quantitative MRI (qMRI) methods, such as measurements of different relaxation times, allow reducing the subjectivity by providing numerical values on which diagnostic assessment or predictions of tissue properties can be based on.

However, such quantitative MRI measurements necessarily take more time than standard anatomical imaging. For example, in \trho{} mapping \cite{gilani2016,borthakur2006,}, typically 5-7 sets of measurements with varying spin lock times are collected to estimate the \trho{} map. Such measurements will thus take 5-7 times longer than acquiring similar anatomical images, often approaching 10 minutes for a stack of quantitative 2D-images. 

\trho{} imaging is based on tilting the magnetization into the $xy$-plane and then locking the magnetization with a spin-lock pulse of certain amplitude and duration. Quantitative mapping, i.e. the measurement of the \trho{} relaxation time constant, is realized by repeating the \trho{} preparation with several different durations of the spin-lock pulse, and collecting the full MR image for each of these preparations. The \trho{} MRI contrast is particularly sensitive to molecular processes occurring at the frequency ($\omega_1$) of the spin-lock pulse, corresponding to the amplitude of the pulse: $\omega_1 = \gamma B_1$, where $\gamma$ is the gyromagnetic ratio, which ties the magnetic field strength (of the RF pulse) $B_1$ to its resonance frequency. Generally, spin-lock pulses operate at and are limited to frequencies that correspond to slow molecular processes that are often both biologically important and altered in disease-related changes. The \trho{} relaxation time has been reported as a promising biomarker for numerous tissues and diseases, such as different disorders of the brain \cite{kettunen2000,grohn2000}, cardiomyopathy \cite{wangc2015}, liver fibrosis \cite{wangyx2011}, musculoskeletal disorders \cite{borthakur2006,wangl2015,kajabi2021} and many others. For a broader overview of \trho{} relaxation and its applications, the reader is referred to the reviews by Gilani and Sepponen \cite{gilani2016}, Wang and Regatte \cite{wangl2015} and Borthakur et al \cite{borthakur2006}.

Staying still in the scanner for extended periods of time can prove to be challenging, for example, for pediatric patients. The excessively long data acquisition times are also operationally impractical because they lead to a small number of studies that can be performed daily with a single MRI device. Quantitative MRI and \trho{} imaging in particular can thus greatly benefit from using undersampled measurements, which are a natural and efficient way to reduce the scanning time for a single qMRI experiment. When using undersampled data, conventional MR image reconstruction methods, such as regridding \cite{Jac+91}, may lead to insufficient reconstruction quality. The usage of compressed sensing (CS) \cite{CRT06,Don06} methods, where an iterative reconstruction method is used together with a sparsifying transform of the image, has proven highly successful with undersampled MRI data \cite{JFL15}.

Usage of CS methods for \trho{} imaging have been previously studied, for example, in \cite{Zhu+15,Pan+16,Zib+18,Zib+20}. In \cite{Zhu+15}, the authors used principal component analysis and dictionary learning in the first in-vivo application of CS to \trho{} reconstruction. In \cite{Pan+16}, the authors used spatial TV together with Autocalibrating Reconstruction for Cartesian sampling (ARC) to accelerate the measurements. In \cite{Zib+18}, the authors compared 12 different sparsifying transforms in 3D-\trho{} mapping. The regularization model combining spatial TV with 2nd order contrast TV was found to perform the best, with satisfactory results with acceleration factor (AF, i.e., the number of datapoints in full data divided by the number of data used in the reconstruction) up to 10 when using cartesian 3D sampling together with parallel imaging. In \cite{Zib+20}, both cartesian and radial data were reconstructed using various different regularization methods. The authors reached acceptable accuracy with AF up to 4 for the cartesian data, whereas with the radial data, the accuracy was acceptable with AF up to 10.

When using CS for \trho{} mapping, the image series with varying spin-lock durations $T_\mathrm{SL}$ is first reconstructed, followed by a pixel-by-pixel non-linear least squares fit of a monoexponential (or a biexponential) signal model to the reconstructed image intensity data to obtain the desired \trho{} relaxation time map. Since the exponential signal model combining the \trho{} and varying $T_\mathrm{SL}$ is well known, a direct, embedded model may also be used to reconstruct the desired \trho{} map directly from the k-space measurement data without the intermediate step of reconstructing the separate intensity maps corresponding to different $T_\mathrm{SL}$.

The direct one-step reconstruction utilizing the embedded model has clear advantages over the sequential two step reconstruction model. First, it reduces the number of unknowns in the reconstruction problem significantly; for example, for measurements with 7 spin-lock times, the number of unknowns may be reduced from $14N$ (one complex image for each contrast) to just $3N$ (\trho{}, \szero{} and a single phase map), where $N$ is the number of pixels or voxels in a single image. Secondly, it allows regularization of the parameter map of interest, i.e., the \trho{} parameter map in case of \trho{} mapping, instead of posing regularization on the complex-valued images corresponding to different contrasts in the intermediate step. And thirdly, since the signal model is embedded in the reconstruction, there is no need to decide what type of a contrast regularization model fits the data best. A disadvantage of the embedded model is that it transforms the MRI inversion into a non-linear problem. The resulting non-linear optimization problem can, however, be solved conveniently with, for example, the non-linear primal-dual proximal splitting algorithm \cite{Val14}.

In this work, we propose an embedded parameterization model to directly reconstruct the \trho{}, \szero{}, and phase maps from the k-space measurement data, and use the non-linear primal-dual proximal splitting algorithm to solve the problem. The proposed model is tested with 2D simulated radial phantom data and 2D cartesian \emph{ex vivo} mouse kidney data. The proposed embedded model is compared with two CS models: one with spatial TV and TV over the $T_\mathrm{SL}$ contrasts, which, we believe, is generally the most commonly used CS model in MRI, and a second CS model with spatial TV and second order contrast TV, which in \cite{Zib+18} was found to perform the best out of 12 different CS models for \trho{} mapping. The first CS model is labeled "CS S1+C1" and the second CS model is labeled "CS S1C2" throughout the paper. The models are named slightly different, since in the first model, the spatial and contrast TV components are separate with two different regularization parameters, and in the second model the spatial TV and the second order contrast TV are under the same root with a single regularization parameter. In the cartesian test case, results from a direct iFFT model are also shown as reference. Reconstructions from both the CS models and the iFFT model are followed by the mono-exponential pixel-by-pixel \trho{} fit.

%%%%%%%%%%%%%%%%%%%%%%%%%%%%%%%%%%%%%%%%%%%%%%%%%%%%%%%%%%%%%%%%%%
\section{Reconstruction methods}
\subsection[Embedded T1rho model]{Embedded \trho{} model}\label{ssec:embeddedModel}
The measurement model for embedded mono-exponential \trho{} mapping is
\begin{equation}
m = K(\szeroTwo,\trhoTwo,\theta)+e,
\end{equation}
where the vectors \szero{}, \trho{}, and $\theta\in\mathbb{R}^N$ are the parameter maps to be reconstructed. The complex measurement vector $m\in\mathbb{C}^{CM}$ is composed of $C$ contrasts each consisting of $M$ k-space measurements. Further, we denote the complex-valued measurement noise by $e\in\mathbb{C}^{CM}$, and the non-linear forward model by $K: \mathbb{R}^{3N} \to \mathbb{C}^{CM}$.

The non-linear forward model can be further decomposed to
\begin{equation}
K(\szeroTwo,\trhoTwo,\theta) = AB(D(\szeroTwo,\trhoTwo,\theta)),
\end{equation}
where $A$ is the block-diagonal matrix containing the Fourier transform operations. In case of cartesian measurements, the blocks of $A$ read $A^c=U^c\mathcal{F}$, where $U^c$ is the undersampling pattern used with the measurements with contrast index $c$, and $\mathcal{F}$ is the Fourier transform. In case of non-cartesian measurements, we approximate the forward model using the non-uniform fast Fourier transform (NUFFT \cite{FS03}), i.e., $A^c=P^c\mathcal{F}L^c$, where $P^c$ is an interpolation and sampling matrix, and $L^c$ is a scaling matrix. Furthermore, $D$ maps the \szero{} and \trho{} parameter maps to magnitude images as
\begin{align}
D(\szeroTwo,\trhoTwo,\theta)=&\left[\begin{array}{c}
\diag(\szeroTwo)\cdot \diag(\exp(-T_{\mathrm{SL}_1}/\trhoTwo))\\
\vdots\\
\diag(\szeroTwo)\cdot\diag(\exp(-T_{\mathrm{SL}_C}/\trhoTwo))\\
\diag(\theta)
\end{array}\right]\\
:=& \left[\begin{array}{c}
\diag(r_1)\\
\vdots\\
\diag(r_C)\\
\diag(\theta)
\end{array}\right],
\end{align}
where $\exp$ denotes elementwise exponentiation, and the division of the scalars $T_{\mathrm{SL}_i}$ by the vector \trho{} is likewise done elementwise. Moreover, $B$ maps the magnitude and phase components of the images to real and complex components, and can be expressed as
\begin{equation}
B(r_1,...,r_C,\theta)=\left[\begin{array}{c}
\diag(r_1)\cdot\diag(\cos\theta)\\
\diag(r_1)\cdot\diag(\sin\theta)\\
\vdots\\
\diag(r_C)\cdot\diag(\cos\theta)\\
\diag(r_C)\cdot\diag(\sin\theta)
\end{array}\right].
\end{equation}
Here too, the $\sin$ and $\cos$ are to be interpreted as elementwise operations. Note that if the phase maps vary between contrasts, the model can be easily modified to reconstruct separate phase maps for all contrasts instead of reconstructing only a single phase map. 
In a typical \trho{} measurement, however, the contrast preparation is usually well-separated from the imaging segment and thus the phase can be expected to be the same between the otherwise identical image acquisition segments.

In the embedded reconstruction, we use total variation regularization for the \szero{} and \trho{} maps, and L2-norm regularization for the spatial gradient of the phase map. We also limit the \szero{} and \trho{} parameter maps above a small positive value. With these, the minimization problem reads
\begin{multline}
\min_{\szeroTwo,\trhoTwo,\theta}||K(\szeroTwo,\trhoTwo,\theta)-m||_2^2 +\alpha_1 \mathrm{TV_S}(\szeroTwo) +\alpha_2 \mathrm{TV_S}(\trhoTwo) \\ +\alpha_3 ||\nabla_\mathrm{S} \theta||_2^2
+\delta_\mathrm{a_1}(\szeroTwo)+\delta_\mathrm{a_2}(\trhoTwo)
\label{eq:emb_t1rho_model}
\end{multline}
where $\alpha_1$, $\alpha_2$, and $\alpha_3$ are the regularization parameters for the \szero{}, \trho{}, and phase maps, respectively, and $a_1$ and $a_2$ are the small positive constraints on the \szero{} and \trho{} maps, respectively.

%%%%%%%%%%%%%%%%%%%%%%%%%%%%%%%%%%%%%%%%%%%%%%%%%%%%%%%%%%%%%%%%%%
\subsubsection[Solving the embedded T1rho reconstruction problem]{Solving the embedded \trho{} reconstruction problem}
The non-linear, non-smooth optimization problem \eqref{eq:emb_t1rho_model} is solved using the non-linear primal-dual proximal splitting algorithm proposed in \cite{Val14}, which is described in Algorithm~\ref{alg:nl-pdps} in its most general form. The algorithm applied to the embedded \trho{} reconstruction is described in more detail in the Appendix.

\begin{algorithm}
	\caption{Non-linear primal-dual proximal splitting \cite[Algorithm 2.1]{Val14}}
	\label{alg:nl-pdps}
	\begin{algorithmic}
		\State $\mathrm{Choose}\ \omega\geq 0\ \mathrm{, and}\ \tau,\sigma$
		\State $\mathrm{s.t.}\ \tau\sigma (\sup_{k=1,...,i}\norm{\nabla H(x^k)}^2)<1.$
		\While{Not reached stopping criterion}
		\State $x^{i+1} := (I+\tau\partial G)^{-1}(x^i-\tau[\nabla H(x^i)]^*y^i)$
		\State $\bar{x}^{i+1} := x^{i+1} + \omega(x^{i+1}-x^i)$
		\State $y^{i+1} := (I+\sigma\partial F^*)^{-1}(y^i+\sigma H(\bar{x}^{i+1}))$
		\EndWhile
	\end{algorithmic}
\end{algorithm}
In our implementation, $x=(\szeroTwo^\mathrm{T}, \trhoTwo^\mathrm{T}, \theta^\mathrm{T})^\mathrm{T}$, and we initialize the \szero{} and phase parts of $x^0$ using iFFT or adjoint of NUFFT of the $T_\mathrm{SL}=0$ measurements. \trho{} was initialized to a constant value of 20, and the dual variable $y$ was initialized to 0. The non-linear mapping $H:\mathbb{R}^{3N}\to \mathbb{C}^{CM+6N}$ contains the non-linear forward model $K$ and the discrete difference matrices.

In addition, we use varying primal step sizes for the different blocks of the embedded reconstruction, i.e. different $\tau_i$ parameters for the \szero{}, \trho{}, and phase updates \cite{MJV20}. This essentially replaces the scalar step length parameter $\tau$ in Algorithm~\ref{alg:nl-pdps} with the diagonal matrix
\begin{equation}
\Tau = \left[\begin{array}{ccc}
\tau_1 I_N & 0 & 0\\
0 & \tau_2 I_N & 0\\
0 & 0 & \tau_3 I_N\\
\end{array}\right].
\end{equation}
The step parameters $\tau_1$, $\tau_2$, and $\tau_3$ are derived from the norm of the corresponding block of the matrix $\nabla K$. Here, we only use the non-linear part $K$ of $H$ to estimate the step lengths, as the linear part of $H$ has only a minor impact on the norm of $\nabla H$. We set the parameter $\sigma$ to $\sigma=1/\max(\tau_i)$ and use $\omega=1$ for the relaxation parameter.

Since the block-diagonal matrix $A$ is linear and can be normalized to 1, we have $\norm{\nabla K} = \norm{J_BJ_D}$. Furthermore, the product of the Jacobians writes
\begin{multline}
\label{eq:J_BJ_D}
J_BJ_D =\\
\left[ \begin{smallmatrix}
\diag(\cos\theta) E_1 & \diag((T_\mathrm{SL_1}/(\trhoTwo\odot\trhoTwo))\odot\cos\theta) r_1 & -\diag(\sin\theta)r_1 \\
\diag(\sin\theta) E_1 & \diag((T_\mathrm{SL_1}/(\trhoTwo\odot\trhoTwo))\odot\sin\theta) r_1 & \diag(\cos\theta)r_1 \\
\diag(\cos\theta) E_2 & \diag((T_\mathrm{SL_2}/(\trhoTwo\odot\trhoTwo))\odot\cos\theta) r_2 & -\diag(\sin\theta)r_2 \\
\diag(\sin\theta) E_2 & \diag((T_\mathrm{SL_2}/(\trhoTwo\odot\trhoTwo))\odot\sin\theta) r_2 & \diag(\cos\theta)r_2 \\
\vdots & \vdots & \vdots \\
\diag(\cos\theta) E_C & \diag((T_\mathrm{SL_C}/(\trhoTwo\odot\trhoTwo))\odot\cos\theta) r_C & -\diag(\sin\theta)r_C \\
\diag(\sin\theta) E_C & \diag((T_\mathrm{SL_C}/(\trhoTwo\odot\trhoTwo))\odot\sin\theta) r_C & \diag(\cos\theta)r_C \\
\end{smallmatrix}\right],
\\ \
\end{multline}
where $E_i=\diag(\exp(-T_\mathrm{SL_i}/\trhoTwo))$, $r_i=\diag(\szeroTwo)\cdot E_i$, and $\odot$ is the Hadamard product, i.e., elementwise multiplication. Now, since the matrix $J_BJ_D$ consists of only diagonal blocks, and the index of the maximum value is the same for all $E_i$, it is straightforward to estimate the $\tau_i$ from the norms of the maximum values of the column-blocks of \eqref{eq:J_BJ_D} yielding
\begin{align}
\tau_1 =& \sqrt{{\sum_{i=1}^C \norm{\exp(-T_\mathrm{SL_i}/\trhoTwo)}_\infty^2}} \label{eq:tau_1}\\
\tau_2 =& \sqrt{\sum_{i=1}^C \norm{r_i(T_\mathrm{SL_i}/(\trhoTwo\odot\trhoTwo))}_\infty^2}\label{eq:tau_2}\\
\tau_3 =& \sqrt{\sum_{i=1}^C \norm{\szeroTwo\odot\exp(-T_\mathrm{SL_i}/\trhoTwo)}_\infty^2} \label{eq:tau_3}.
\end{align}

In addition, we calculate the norms in every iteration, and update the used $\tau_i$ and $\sigma$ if the step is smaller than the previously used step.

In our experience, these step lengths may, however, prove to be too small, and in some cases larger step lengths, especially for the \trho{} update step, may be used to obtain faster convergence. In this work, we used a multiplier of 50 for the \trho{} update step $\tau_2$ in the radial simulation. Note that the step length criterion of Algorithm~\ref{alg:nl-pdps} still holds with the multiplier, since $\tau_2\cdot\sigma$ remains small due to the selection of $\sigma$.

%%%%%%%%%%%%%%%%%%%%%%%%%%%%%%%%%%%%%%%%%%%%%%%%%%%%%%%%%%%%%%%%%%
\subsection{Compressed sensing reference methods}

We compare the embedded model to two CS models, which include a complex valued reconstruction of the images with different spin-lock times, followed by a pixel by pixel non-linear least squares fit of the monoexponential signal model to obtain the \trho{} and \szero{} parameter maps. The first CS reconstruction model uses spatial total variation together with first order total variation over the varying $T_{SL}$ contrasts (labeled CS S1+C1) and the second one uses spatial total variation together with second order total variation over the varying $T_{SL}$ contrasts (labeled CS S1C2).

The measurement model for a single contrast image is
\begin{equation}
\label{eq:cs_meas_model}
m^c  =A^c u^c+e^c,
\end{equation}
where the superscript $c$ denotes the contrast index, $m^c\in\mathbb{C}^M$ is the k-space data vector for contrast index $c$, $u^c\in\mathbb{C}^N$ is the image vector, $e^c\in\mathbb{C}^M$ is the complex valued noise vector, and $A^c$ is the forward model, which depends on the measurement sequence and undersampling pattern, and is described in more detail in Section~\ref{ssec:embeddedModel}. 

With the measurement model of \eqref{eq:cs_meas_model}, spatial total variation, and total variation over the contrasts, the CS minimization problem reads
\begin{equation}
u^*=\argmin_{u} \norm{Au-m}_2^2+ \alpha \mathrm{TV_S}(u) + \beta \mathrm{TV_C}(u),
\label{eq:cs_model}
\end{equation}
where $A$ is a block-diagonal matrix containing the forward transforms $A^c$ corresponding to each image, $u\in\mathbb{C}^{NC}$ is all the images vectorized, such that $C$ is the number of contrasts, and $m\in\mathbb{C}^{MC}$ is all the k-space measurements vectorized. Further, $\mathrm{TV_S}$ denotes spatial total variation, $\mathrm{TV_C}$ denotes total variation over contrasts, and $\alpha$ and $\beta$ are the regularization parameters of spatial and contrast TV respectively.

The second CS minimization problem, which uses the single regularization parameter version of combined spatial TV and second order contrast TV, reads
\begin{equation}
u^*=\argmin_{u} \norm{Au-m}_2^2+ \alpha \mathrm{TV_{SC}(u)},
\label{eq:cs_s1c2_model}
\end{equation}
where 
\begin{equation*}
\mathrm{TV_{SC}}(u)=\sum_k{\sqrt{(\nabla_\mathrm{x} u)_k^2+(\nabla_\mathrm{y} u)_k^2+(\nabla_\mathrm{c}^2 u)_k^2 }},
\end{equation*}
where $\nabla_\mathrm{x}$ and $\nabla_\mathrm{y}$ are the horizontal and vertical direction spatial discrete forward difference operators respectively, $\nabla_\mathrm{c}^2$ is the second order contrast direction discrete difference operator, and $k$ is an index that goes through all the pixels in the set of images.

Both of the minimization problems (\ref{eq:cs_model}, \ref{eq:cs_s1c2_model}) are solved using the popular primal-dual proximal splitting algorithm of Chambolle and Pock \cite{CP11}.

Finally, in the CS models (and the iFFT model), we fit the mono-exponential \trho{} signal equation
\begin{equation}
\label{eq:t1rhomodel}
[\mathrm{T}_{\mathrm{1\rho},k}^*,\mathrm{S}_{0,k}^*]=\argmin_{\mathrm{T}_\mathrm{1\rho},\mathrm{S}_0} \norm{|u_k|-\mathrm{S}_0\exp(-T_\mathrm{SL}/\mathrm{T}_{1\rho})}^2
\end{equation}
pixel by pixel to the reconstructed intensity images obtained by solving either \eqref{eq:cs_model} or \eqref{eq:cs_s1c2_model}. Here $|u_k|=|u_k^1|,...,|u_k^C|$ is the vector of reconstruction intensity values at pixel location $k$ with $T_\mathrm{SL}$ contrasts 1 to $C$, and similarly $T_\mathrm{SL}$ is the vector of $T_\mathrm{SL}$ values of contrasts 1 to $C$. Note that the final \szero{} estimate is obtained from the mono-exponential model fit instead of taking the intensity values from the reconstructions with $T_\mathrm{SL}=0$.

%%%%%%%%%%%%%%%%%%%%%%%%%%%%%%%%%%%%%%%%%%%%%%%%%%%%%%%%%%%%%%%%%%
\section{Materials and Methods}
\subsection{Simulated golden angle radial data}
\begin{figure}
	\centering
	\includegraphics[width=\columnwidth]{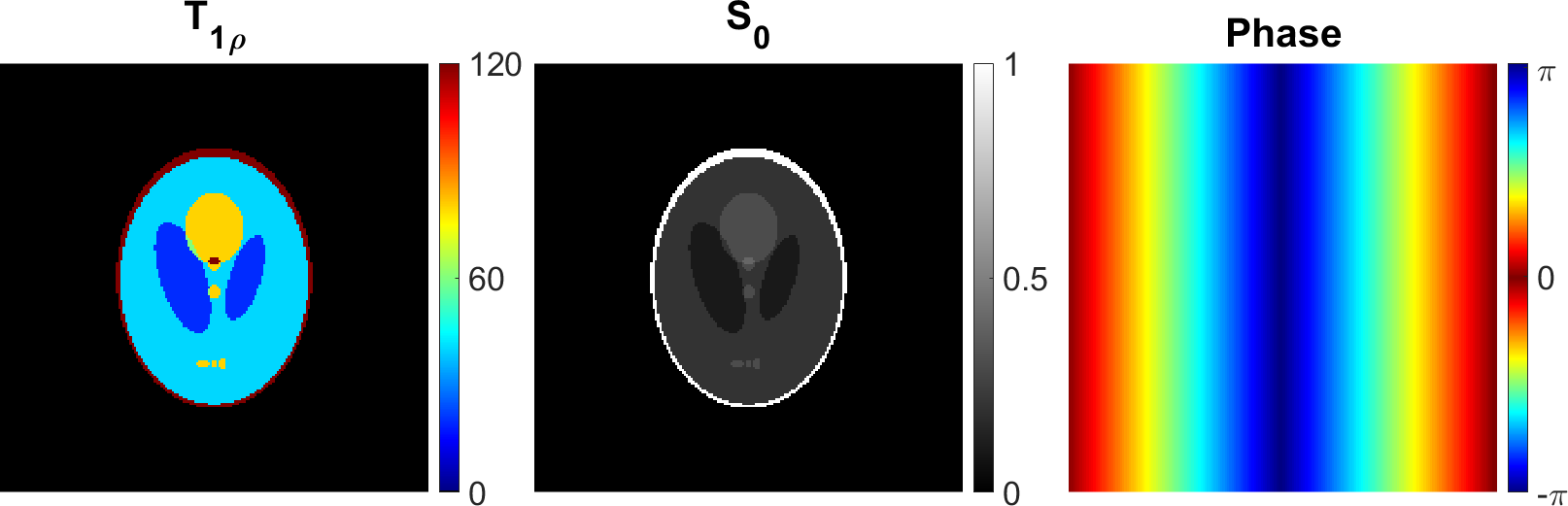}
	\caption{The simulated \trho{}, \szero{}, and phase parameter maps for the radial golden angle \trho{} phantom simulation.}
	\label{fig:simu_simuParts}
\end{figure}

The simulation of the radial measurement data was based on the Shepp-Logan phantom in dimensions $128\cdot 128$, which was zero-filled to dimensions $192\cdot 192$. The \trho{} values of the target were set to between 20 and 120. The intensity with $T_\mathrm{SL}=0$ was set to a maximum of 1, and the phase of the target was set $2\pi x/192$, where $x$ is the horizontal coordinate of the pixel. The images of the simulated \trho{}, \szero{}, and phase maps are shown in Figure~\ref{fig:simu_simuParts}. To generate the varying $T_\mathrm{SL}$ measurements, spin lock times of 0, 4, 8, 16, 32, 64, and 128 ms were used. For each $T_\mathrm{SL}$, 302 (i.e., $\sim192\cdot\pi/2$) golden angle \cite{Win+07} spokes were generated. This corresponds to full sampling for equispaced radial spokes with image dimensions $192\cdot 192$ in the sense that the distance between spokes at their outermost points satisfies the Nyquist criterion \cite{BKZ04}. Finally, complex Gaussian noise at 5 \% of the mean of the absolute values of the full noiseless simulation was added to the simulated measurements.

{}

%%%%%%%%%%%%%%%%%%%%%%%%%%%%%%%%%%%%%%%%%%%%%%%%%%%%%%%%%%%%%%%%%%
\subsection{Cartesian data from ex vivo mouse kidney}
Experimental \emph{ex vivo} data from a mouse kidney was acquired from a separate study. The data was collected in compliance with ethical permits (ESAVI-2020-006283) at 9.4 T using a 19 mm quadrature RF volume transceiver (RAPID Biomedical GmbH, Rimpar, Germany) and VnmrJ3.1 Varian/Agilent DirectDrive console. \trho{} relaxation data was collected using a refocused \trho{} preparation scheme \cite{witschey2007} with a spin-lock frequency of 500 Hz and $T_\mathrm{SL}$ = 0, 8, 16, 32, 64, and 128 ms. The \trho{}-prepared data, i.e., \trho{}-weighted images, were collected using a fast spin echo sequence with a repetition time of 5 s, effective echo time of 5.5 ms, echo train length of 8, slice thickness of 1mm, field-of-view of 17 x 17 mm and acquisition matrix of 192 x 192. Eventually, only spin-lock times up to 64 ms were used in the reconstruction as the signal intensity of the longest spin-lock time was close to the noise level and had minimal or no effect on the reconstruction.

%%%%%%%%%%%%%%%%%%%%%%%%%%%%%%%%%%%%%%%%%%%%%%%%%%%%%%%%%%%%%%%%%%
\subsection{Reconstruction specifics}
The radial data from the 2D phantom was reconstructed with the embedded model and the two CS models with acceleration factors 1, 5, 10, 20, 30, 50, and 101 (rounded to the nearest integer). In \trho{} imaging, the images measured with varying spin-lock times are expected to have high redundancy in the sense that the images are expected to be structurally similar with decreasing intensity as $T_\mathrm{SL}$ increases, making complementary k-space sampling warranted. In complementary k-space sampling, the subsampling with any measured contrast is different from the others, meaning that each sampling adds to spatial information gained at other contrasts. The golden angle radial sampling is especially well suited for this as the measurements are inherently complementary (i.e., each new spoke has a different path in the k-space compared to the previous ones) and each measured spoke traverses through the central (low-frequency) part of the k-space which contains significant information on the intensity level of  the  images. Thus, we sampled the golden angle data such that, for example, with an acceleration factor 10, the first contrast used the first 30 spokes out of the 302 total, the second contrast used spokes 31 through 60 and so on to achieve complementary k-space sampling.

In the embedded model, the phase regularization parameter was set to a constant value at 0.01, and the other regularization parameters were varied over a wide range. In the CS models, the regularization parameters were also varied over a wide range to find the best parameters. The reconstructions shown use the regularization parameters that yielded the smallest \trho{} RMSE with respect to the ground truth phantom.

The cartesian \emph{ex vivo} mouse kidney data was reconstructed with the embedded, the iFFT, and the two CS methods with acceleration factors 2, 3, 4, and 5 (rounded to the nearest integer). Undersampling was done by taking a number of full k-space rows corresponding to the desired acceleration factor, since cartesian data collection in MRI scanners is done line by line. For the undersampled reconstructions, 1/4 of the total sampled k-space rows were taken from around the center to include zero frequency and enough low frequency data in all contrasts. Half of the rest 3/4 were taken from the top part and the other half from the bottom part. To achieve complementary sampling, the rows from the top and bottom parts were selected such that all rows were first selected once in random order, before continuing to sample from the full set of rows again.

In the \emph{ex vivo} test case, too, the phase regularization parameter of the embedded model was set to a constant level, which was 0.0001, and the other parameters of the embedded and both CS models were varied over a wide range to find the optimal \trho{} estimate. The embedded model reconstructions were compared to the embedded reconstruction with full data and likewise the CS and iFFT model reconstructions were compared to the corresponding reconstructions with full data as the true \trho{} map is not available. Thus, the RMSEs reflect each model's relative tolerance for undersampling compared to the situation where fully sampled data are available for the particular reconstruction model.

\ % Text color bug without this empty line

%%%%%%%%%%%%%%%%%%%%%%%%%%%%%%%%%%%%%%%%%%%%%%%%%%%%%%%%%%%%%%%%%%
\section{Results}
\subsection{Simulated golden angle radial data}
\begin{figure}
	\centering
	\includegraphics[width=\columnwidth]{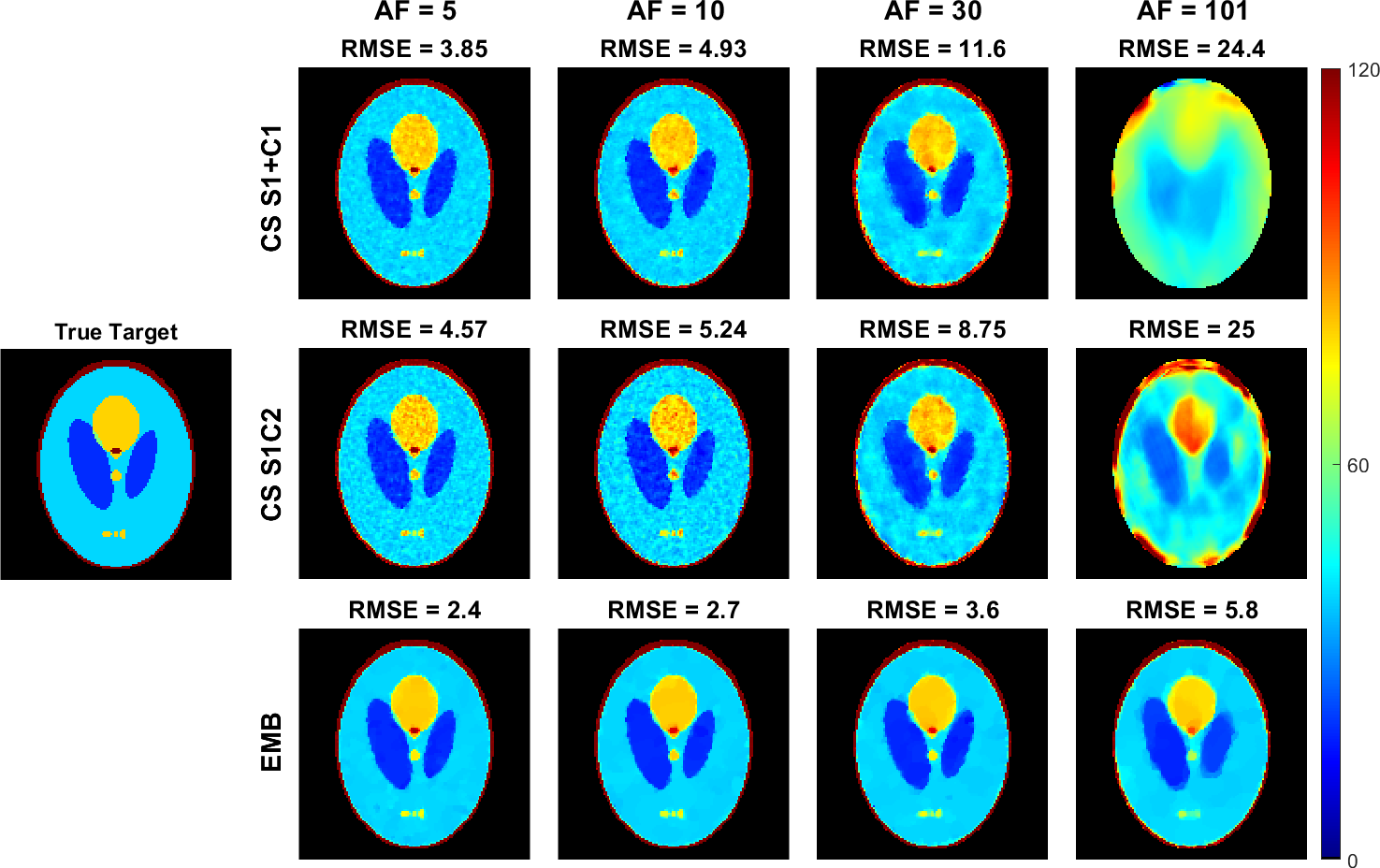}
	\caption{The \trho{} maps of the radial simulation reconstructed with the embedded model and the two CS models, and the RMSEs of the reconstructions as compared to the true values used in the simulation. The top row contains the CS S1+C1 model, and the middle row the CS S1C2 model \trho{} parameter maps obtained from the monoexponential fit of \eqref{eq:t1rhomodel}, and the bottom row contains the embedded model reconstructions. Columns 2-5 show the \trho{} parameter maps at acceleration factors 5, 10, 30, and 101. Images are cropped to content.}
	\label{fig:simu_t1rho}
\end{figure}

\begin{figure}
	\centering
	\includegraphics[width=\columnwidth]{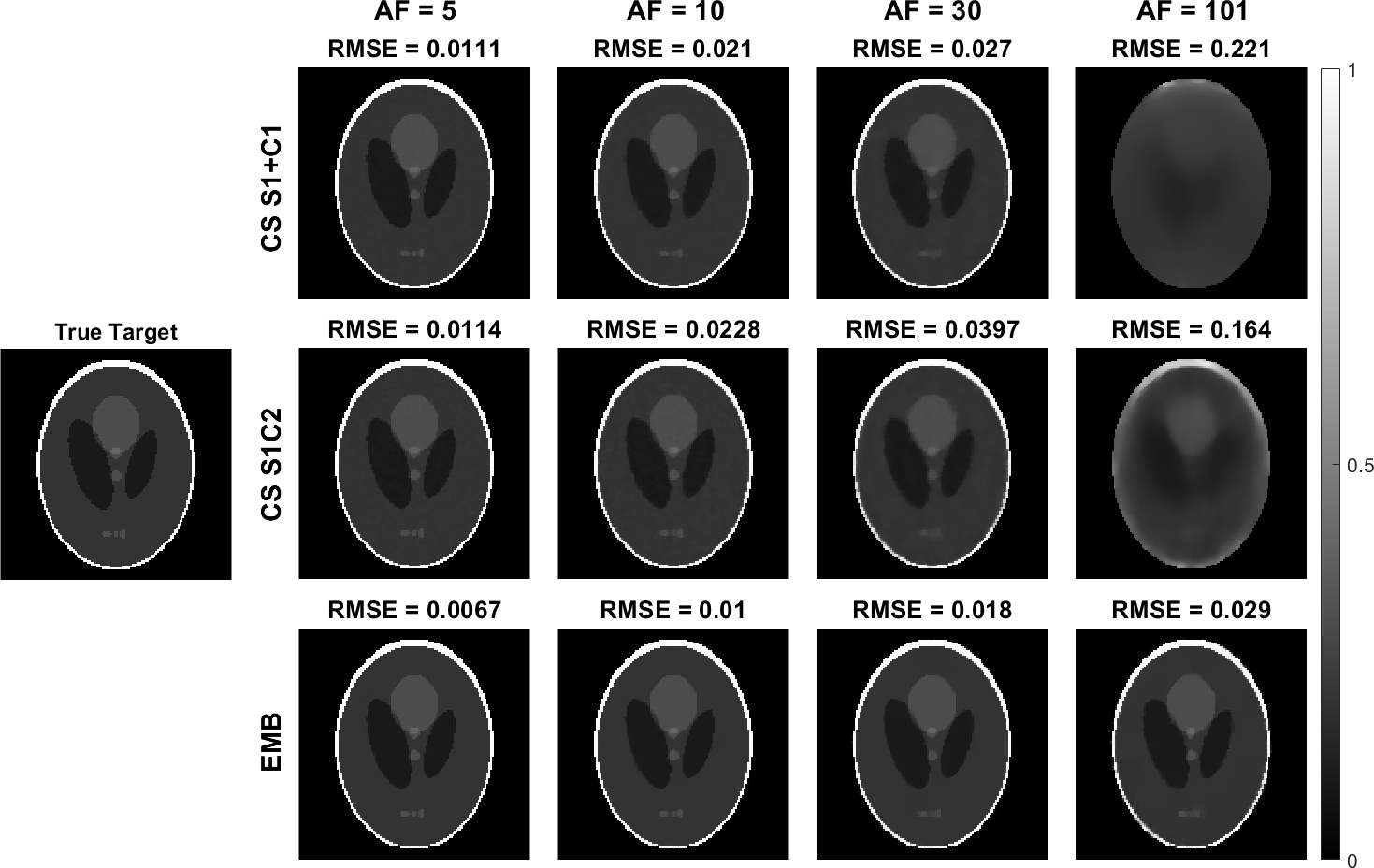}
	\caption{The \szero{} maps of the radial simulation reconstructed with the embedded model and the two CS models, and the RMSEs of the reconstructions as compared to the true values used in the simulation. The \szero{} maps shown here are from the same reconstructions as the \trho{} maps shown in Figure~\ref{fig:simu_t1rho}. The top row contains the CS S1+C1 model, and the middle row the CS S1C2 model \szero{} parameter maps obtained from the monoexponential fit of \eqref{eq:t1rhomodel}, and the bottom row contains the embedded model reconstructions. Columns 2-5 show the \szero{} parameter maps at acceleration factors 5, 10, 30, and 101. Images are cropped to content.}
	\label{fig:simu_s0}
\end{figure}

With the radial simulated phantom data, all the methods produce reconstructions with similar RMSE when using full data (acceleration factor 1). With undersampled data, the embedded model outperforms both the CS models as measured by RMSE of both the \trho{} (Figure~\ref{fig:simu_t1rho}) and \szero{} (Figure~\ref{fig:simu_s0}) maps with all acceleration factors, and the improvement increases with larger acceleration factors.

{}

{}

\begin{figure}
	\centering
	\includegraphics[width=\columnwidth]{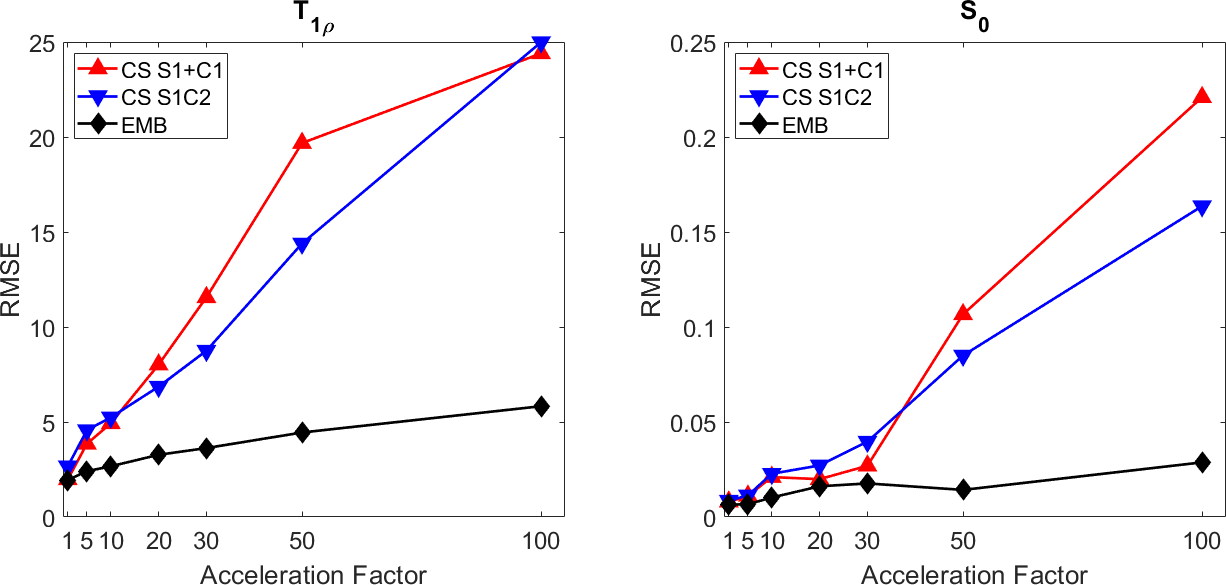}
	\caption{The RMSEs of the \trho{} (left) and \szero{} (right) maps of the radial simulation with the embedded model and the two CS models at acceleration factors 1, 5, 10, 20, 30, 50, and 101.}
	\label{fig:simu_rmse}
\end{figure}{}

The \trho{} maps computed using the CS models are also visibly noisier as the model does not allow direct regularization of the \trho{} map (Figure~\ref{fig:simu_t1rho}). With an acceleration factor of 101, reconstructions of both CS models start to break down, whereas the embedded model reconstruction still reconstructs the target reasonably well, with the RMSE values below those of the the CS models at an acceleration factor of 20--30 (Figures~\ref{fig:simu_t1rho},~\ref{fig:simu_s0} and~\ref{fig:simu_rmse}).

%%%%%%%%%%%%%%%%%%%%%%%%%%%%%%%%%%%%%%%%%%%%%%%%%%%%%%%%%%%%%%%%%%
\subsection{Cartesian data from ex vivo mouse kidney}
\begin{figure}
	\centering
	\includegraphics[width=\columnwidth]{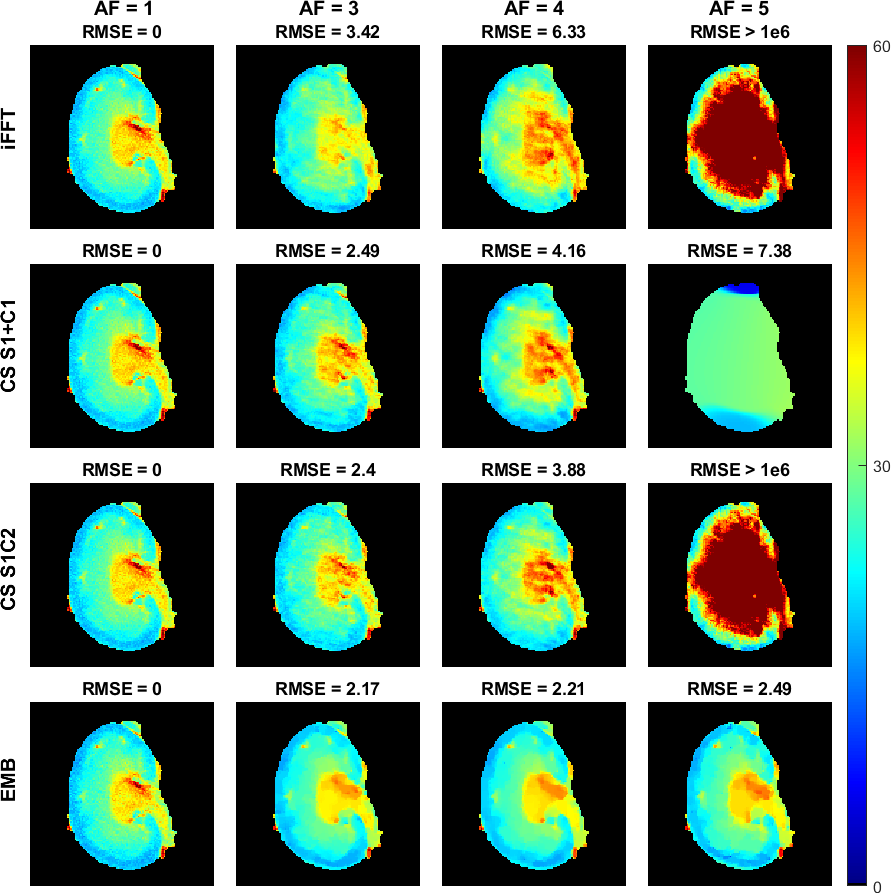}
	\caption{The \trho{} maps of the cartesian \emph{ex vivo} mouse kidney data with the iFFT, CS S1+C1, CS S1C2 and embedded models, and the RMSEs as compared to the corresponding model reconstructions with full data. The top row contains the iFFT, the second row the CS S1+C1, and the third row the CS S1C2 model \trho{} parameter maps obtained from the monoexponential fit of \eqref{eq:t1rhomodel}, and the bottom row contains the \trho{} maps obtained from the embedded model reconstructions. Columns 1-4 show the parameter maps corresponding to acceleration factors 1, 3, 4, and 5. Images are cropped to content.}
	\label{fig:expeKidney_t1rho}
\end{figure}

\begin{figure}
	\centering
	\includegraphics[width=\columnwidth]{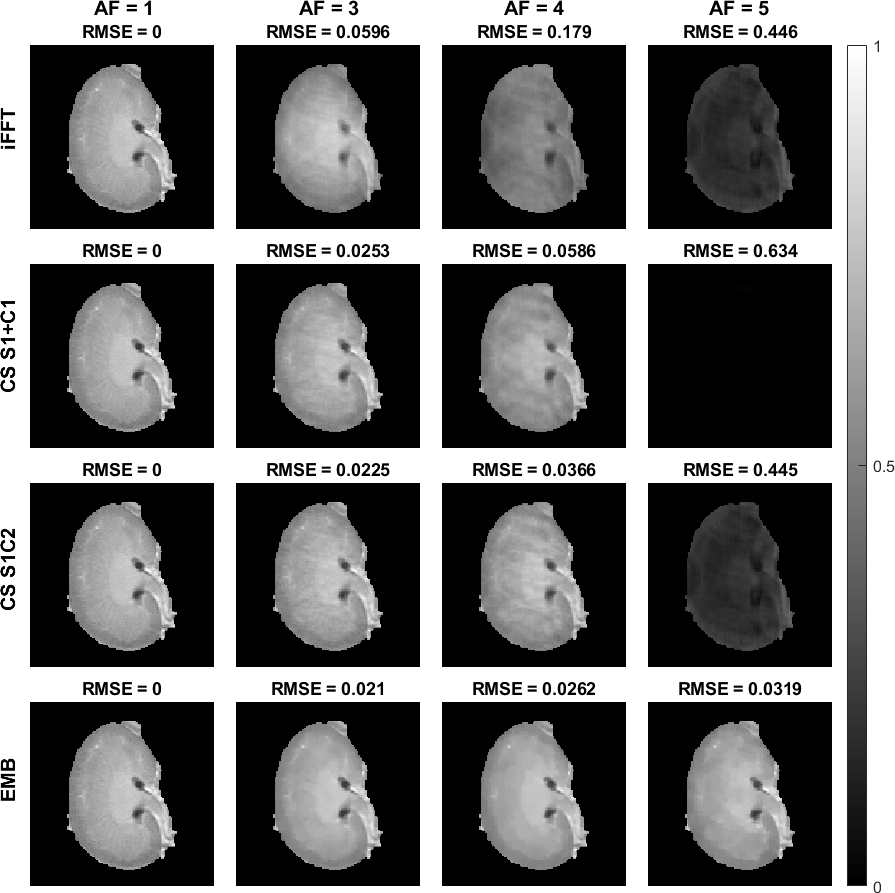}
	\caption{The \szero{} maps of the cartesian \emph{ex vivo} mouse kidney data with the iFFT, CS S1+C1, CS S1C2 and embedded models, and the RMSEs as compared to the corresponding model reconstructions with full data. The \szero{} maps shown here are from the same reconstructions as the \trho{} maps shown in Figure~\ref{fig:expeKidney_t1rho}. The top row contains the iFFT, the second row the CS S1+C1, and the third row the CS S1C2 model \szero{} parameter maps obtained from the monoexponential fit of \eqref{eq:t1rhomodel}, and the bottom row contains the \szero{} maps obtained from the embedded model reconstructions. Columns 1-4 show the parameter maps corresponding to acceleration factors 1, 3, 4, and 5. Images are cropped to content.}
	\label{fig:expeKidney_s0}
\end{figure}

\begin{figure}
	\centering
	\includegraphics[width=\columnwidth]{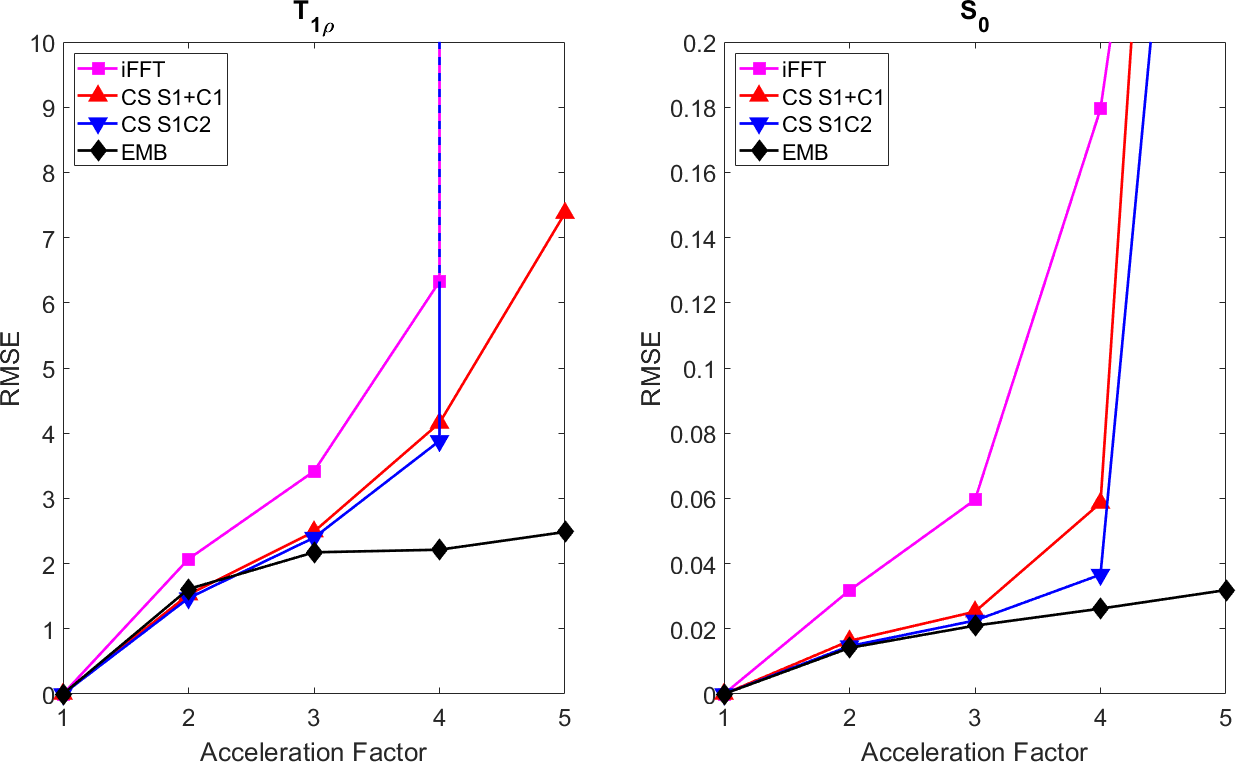}
	\caption{The RMSEs of the \trho{} (left) and \szero{} (right) maps of the cartesian \emph{ex vivo} mouse kidney data with the embedded, CS S1+C1, CS S1C2, and iFFT models at acceleration factors 2, 3, 4, and 5.}
	\label{fig:expeKidney_rmse}
\end{figure}

In the cartesian \emph{ex vivo} test case, the performance of the embedded and CS models in their relative tolerance for undersampling is similar with acceleration factor 2, and both CS models perform slightly worse than the embedded model with acceleration factor 3 (Figures~\ref{fig:expeKidney_t1rho},~\ref{fig:expeKidney_s0} and~\ref{fig:expeKidney_rmse}). With acceleration factor 4, the performance of the CS models is already clearly worse than the performance of the embedded model. While both of the CS models fail in the reconstruction with acceleration factor 5, the embedded model still produces similar tolerance for undersampling as with the smaller acceleration factors. The undersampled iFFT reconstructions shown for reference perform worse than the CS or the embedded model reconstructions with all the acceleration factors.

{}

{}

{}

%%%%%%%%%%%%%%%%%%%%%%%%%%%%%%%%%%%%%%%%%%%%%%%%%%%%%%%%%%%%%%%%%%
\section{Discussion}
In this work, we proposed a non-linear, embedded \trho{} model for direct quantitative \trho{} reconstruction. The model is solved using the non-linear primal-dual proximal splitting algorithm \cite{Val14}. We compared the embedded model reconstructions to two compressed sensing reconstructions followed by a mono-exponential \trho{} fit in a radial simulated test case and a cartesian \emph{ex vivo} test case. In the cartesian test case, we also show results from iFFT reconstructions followed by the \trho{} fit. 

In the simulated test case, where the RMSE metric with respect to the true target image is available, the embedded model outperformed both of the CS models with improvement increasing towards the higher acceleration factors. In the experimental test case with Cartesian ex vivo mouse kidney data, the RMSEs reflect the relative tolerance of the method with respect to the case where the fully sampled data was available for that particular method. In this case, the embedded model and the CS models had similar RMSE for acceleration factor 2, and for higher acceleration factors the embedded model exhibited clearly better tolerance for undersampling, indicating that the embedded model would allow usage of higher acceleration factors than the CS models.

The two CS models perform quite similarly with the second-order contrast TV model CS S1C2 performing slightly better overall than the CS S1+C1 model in the simulated test case. The same observation can be made in the cartesian test case up to acceleration factor 4. In the Cartesian test case the CS S1+C1 model has smaller RMSE than CS S1C2 with acceleration factor 5, but in this case both of the CS models failed to produce useful \trho{} or \szero{} maps. From the practical point of view, the second-order contrast TV model with the implementation described in \cite{Zib+18} is also more convenient than the CS S1+C1 model as it requires selecting only a single regularization parameter.

The embedded model is, however, slower to compute than the CS models. For example, our MATLAB code for the radial simulation data with $\mathrm{AF}=5$ took 104 minutes for the embedded model and 26 minutes for the CS S1+C1 model. For the experimental cartesian data, the difference was bigger: for example for $\mathrm{AF}=2$, the embedded model took 75 minutes to compute, while the CS S1+C1 model converged to stopping criterion in under a minute. The computation times could, however, be shortened, for example, by optimizing the code, running the code on a GPU, and also by loosening the stopping criteria since we ran the iterations to rather strict criteria.

In the radial simulated test case, the embedded model reconstructs the target quite well even with an acceleration factor of 101, using only 3 spokes per $T_{SL}$ contrast, and 21 spokes in the whole reconstruction. In the cartesian test case, the acceleration factors that can be reached are much smaller. Even though the target used in the radial simulation is rather simple, it is evident that the radial sampling pattern, particularly with the golden angle sampling where k-space spokes are complementary and go through the center part of the k-space, allows much higher acceleration factors than a cartesian line-by-line sampling pattern. This is due to the undersampling artefacts in radial sampling (i.e. streaking) being more noise like in the transform domain than the undersampling artefacts that arise in cartesian sampling \cite{LDP07,BUF07}. This finding is aligned with the findings of \cite{Zib+20}.

Testing the proposed embedded model with radial experimental data, \emph{in vivo} data, 3D-data, and parallel imaging data are interesting future works, and our hypothesis is that similar results, where the embedded model outperforms the CS models, are to be expected. In addition, the embedded \trho{} model could be tested with other regularizers, such as total generalized variation \cite{BKP10}, which balances between minimizing the first and second-order differences of the signal. 

As the contrast manipulation scheme of the signal acquisition and the quantitative signal equation are the only major aspects that change between different qMRI contrasts, the proposed method can easily be adapted to fit other qMRI cases as well. Besides other qMRI methods, other aspects where embedded modelling could offer further benefits are \trho{} dispersion imaging \cite{akella2004,keenan2015}, where the data are acquired at multiple spin-locking amplitudes, and reducing RF energy deposition by asymmetric data reduction for the different spin-lock times (i.e. less data for long spin-lock pulses). More generally, shorter scan times may allow for higher spin-lock durations and/or higher amplitude pulses, as the specific absorption rate of RF energy can be minimized via acquiring less data for the most demanding pulses. Alternatively, multi-contrast embedded modelling could offer further avenues for data reduction.

%%%%%%%%%%%%%%%%%%%%%%%%%%%%%%%%%%%%%%%%%%%%%%%%%%%%%%%%%%%%%%%%%%
\section{Conclusions}
In this work, we proposed an embedded \trho{} reconstruction method, which directly reconstructs the \trho{}, \szero{}, and phase maps from the measurement data. The reconstruction method also allows direct regularization of these parameter maps, and thus \emph{a priori} information about the parameter maps may be incorporated into the reconstruction. We also showed that the proposed method outperforms two compressed sensing models in two test cases, especially when using higher acceleration factors.

%%%%%%%%%%%%%%%%%%%%%%%%%%%%%%%%%%%%%%%%%%%%%%%%%%%%%%%%%%%%%%%%%%
\begin{appendices}
	\section{Algorithm details}
	The minimization problem we are solving using NL-PDPS reads
	\begin{multline}
	\min_{\szeroTwo,\trhoTwo,\theta}||K(\szeroTwo,\trhoTwo,\theta)-m||_2^2 +\alpha_1 \mathrm{TV_S}(\szeroTwo) +\alpha_2 \mathrm{TV_S}(\trhoTwo)\\ +\alpha_3 ||\nabla_\mathrm{S} \theta||_2^2
	+\delta_\mathrm{a_1}(\szeroTwo)+\delta_\mathrm{a_2}(\trhoTwo).
	\label{eq:emb_t1rho_model_2}
	\end{multline}
	For the algorithm, the minimization problem is written as
	\begin{equation}
	\min_u F(H(u))+G(u),
	\end{equation}
	where $u=(\szeroTwo^\mathrm{T},\trhoTwo^\mathrm{T},\theta^\mathrm{T})^\mathrm{T}:=(u_1^\mathrm{T},u_2^\mathrm{T},u_3^\mathrm{T})^\mathrm{T}$, and
	\begin{equation}
	H(u)= \left[\begin{array}{c}
	K(u)\\
	\nabla_\mathrm{D} u\\
	\end{array}\right], \mathrm{where\ }
	\nabla_\mathrm{D} = \left[\begin{array}{ccc}
	\nabla_\mathrm{x} & 0 & 0\\
	\nabla_\mathrm{y} & 0 & 0\\
	0 & \nabla_\mathrm{x} & 0\\
	0 & \nabla_\mathrm{y} & 0\\
	0 & 0 & \nabla_\mathrm{x}\\
	0 & 0 & \nabla_\mathrm{y}\\
	\end{array}\right],
	\end{equation}
	where $\nabla_\mathrm{x}$ and $\nabla_\mathrm{y}$ are discrete forward differences in the horizontal and vertical directions. Further, the functional $F$ is divided into 4 parts matching the parts of the minimization problem as
	\begin{align}
	F_1(p_1) =& \frac{1}{2}\norm{p_1-m}_2^2 \\
	F_2(p_2) =& \alpha_1\norm{\abs{p_2}}_1\\
	F_3(p_3) =& \alpha_2\norm{\abs{p_3}}_1\\
	F_4(p_4) =& \alpha_3\norm{p_4}_2^2,
	\end{align}
	and the $p_i$ are obtained by
	\begin{align}
	p_1 =& K(u) \\
	p_2 =& \nabla_\mathrm{S} u_1 := \left[\begin{array}{c}
	\nabla_\mathrm{x} u_1\\
	\nabla_\mathrm{y} u_1\\
	\end{array}\right] \\
	p_3 =& \nabla_\mathrm{S} u_2 := \left[\begin{array}{c}
	\nabla_\mathrm{x} u_2\\
	\nabla_\mathrm{y} u_2\\
	\end{array}\right]\\
	p_4 =& \nabla_\mathrm{S} u_3 := \left[\begin{array}{c}
	\nabla_\mathrm{x} u_3\\
	\nabla_\mathrm{y} u_3\\
	\end{array}\right].
	\end{align}
	Here, the $|p_2|$ and $|p_3|$ denote the isotropic gradient magnitudes, i.e., for example for $|p_2|$, the elements are $(|p_2|)_i=\sqrt{(\nabla_\mathrm{x}\szeroTwo)_i^2+(\nabla_\mathrm{y}\szeroTwo)_i^2}$. Similarly, the functional $G$ has two parts, which read
	\begin{align}
	G_1(u) =& \delta_\mathrm{a_1}(\szeroTwo) \\
	G_2(u) =& \delta_\mathrm{a_2}(\trhoTwo).
	\end{align}
	
	Now, the proximal operators (also called the resolvent operators) of the convex conjugates of $F_i$, i.e., $F_i^*$, and of $G_1$ and $G_2$ read
	\begin{align}
	(I+\sigma\partial F_1^*)^{-1}(v_1) =& \frac{v_1-\sigma m}{1+\sigma} \\
	(I+\sigma\partial F_2^*)^{-1}(v_2) =& \frac{v_2}{\max(1,|v_2|/\alpha_1)}\\
	(I+\sigma\partial F_3^*)^{-1}(v_3) =& \frac{v_3}{\max(1,|v_3|/\alpha_2)}\\
	(I+\sigma\partial F_4^*)^{-1}(v_4) =& \frac{v_4}{\sigma/(2\alpha_3)+1}\\
	(I+\tau\partial G_1)^{-1}(u) =& P_{1,a_1}(u) = \begin{cases}
	u_1,& u_1\geq a_1\\
	a_1,& u_1<a_1
	\end{cases}\\
	(I+\tau\partial G_2)^{-1}(u) =& P_{2,a_2}(u) = \begin{cases}
	u_2,& u_2\geq a_2\\
	a_2,& u_2<a_2
	\end{cases}.
	\end{align}
	With these, we can write algorithm \ref{alg:nl-pdps_precise}. The step lengths $\tau_i$ are chosen by Eqs.~\ref{eq:tau_1}-\ref{eq:tau_3}.
	\begin{algorithm}
		\caption{Embedded \trho{} with NL-PDPS \cite[Algorithm 2.1]{Val14}}
		\label{alg:nl-pdps_precise}
		\begin{algorithmic}
			\State $\mathrm{Choose}\ \omega\geq 0\ \mathrm{, and}\ \tau_\ell,\sigma$
			\State $\mathrm{s.t.}\ \tau_\ell\sigma (\sup_{k=1,...,i}\norm{[\nabla H(x^k)]_\ell}^2)<1.$
			\State $\Tau = \diag(\tau_1 I_N,\ \tau_2 I_N,\ \tau_3 I_N).$
			\While{Not reached stopping criterion}
			%\State $u^{i+1} := P_{2,a_2}(P_{1,a_1}(u^i-\Tau[\nabla L(u^i)]^*[v_1^{i^\mathrm{T}} v_2^{i^\mathrm{T}} v_3^{i^\mathrm{T}} v_4^{i^\mathrm{T}}]^\mathrm{T}))$
			\State $\tilde{u}^{i+1} \leftarrow u^i-\Tau[\nabla L(u^i)]^*[v_1^{i^\mathrm{T}} v_2^{i^\mathrm{T}} v_3^{i^\mathrm{T}} v_4^{i^\mathrm{T}}]^\mathrm{T}$
			\State $u^{i+1} \leftarrow P_{2,a_2}(P_{1,a_1}(\tilde{u}^{i+1}))$
			\State $\bar{u}^{i+1} \leftarrow u^{i+1} + \omega(u^{i+1}-u^i)$
			\State $v_1^{i+1} \leftarrow (I+\sigma\partial F_1^*)^{-1}(v_1^i+\sigma K(\bar{u}_1^{i+1}))$
			\State $v_2^{i+1} \leftarrow (I+\sigma\partial F_2^*)^{-1}(v_2^i+\sigma \nabla_\mathrm{S}\bar{u}_2^{i+1})$
			\State $v_3^{i+1} \leftarrow (I+\sigma\partial F_3^*)^{-1}(v_3^i+\sigma \nabla_\mathrm{S}\bar{u}_3^{i+1})$
			\State $v_4^{i+1} \leftarrow (I+\sigma\partial F_4^*)^{-1}(v_4^i+\sigma \nabla_\mathrm{S}\bar{u}_4^{i+1})$
			\EndWhile
		\end{algorithmic}
	\end{algorithm}

\end{appendices}
\printbibliography

\end{document}